\documentclass{webofc}\usepackage[varg]{txfonts}\usepackage{bm}
\usepackage{color}\usepackage{amsmath}\usepackage{slashed}
\usepackage{booktabs}\usepackage{dcolumn}
\woctitle{QCD@Work 2018}\newcolumntype{d}[1]{D{.}{.}{#1}}
\begin{document}\title{Simplified Bethe--Salpeter Description of
Basic\\Pseudoscalar-Meson Features}\author{Wolfgang Lucha
\inst{1}\fnsep\thanks{\email{Wolfgang.Lucha@oeaw.ac.at}}}
\institute{Institute for High Energy Physics, Austrian Academy of
Sciences, Nikolsdorfergasse 18,\\ A-1050 Vienna, Austria}

\abstract{We assess a description of pseudoscalar mesons as
pseudo-Goldstone bosons by its compatibility with some
Gell-Mann--Oakes--Renner-type relation.}\maketitle

\section{Pseudoscalar Mesons of Goldstone Nature}Goldstone's
theorem necessitates the presence of a massless boson in the
spectrum of physical particles for any spontaneously broken chiral
symmetry of quantum chromodynamics (QCD); these hypothetical
Goldstone bosons are identified with the ground-state pseudoscalar
mesons (pions, kaons, $\eta$). Their finite (but comparatively
small) masses are attributed to the additional explicit breakdown
of the chiral symmetries of QCD enforced by nonvanishing quark
masses.

We analyze the Goldstone-boson nature of the lightest pseudoscalar
mesons by means of a formalism \cite{WLe,WLp,WLpa,WLpb} situated
somewhere between the fully relativistic Bethe--Salpeter
description of bound states \cite{SB}, with several still to be
resolved inherent obstacles, and the latter's extreme
instantaneous limit, represented by its three-dimensional
reduction devised by Salpeter \cite{SE}. A rather promising tool
to judge the merits of such kind of intermediate framework proves
to be, among others, the fulfilment of (generalized)
Gell-Mann--Oakes--Renner-type relations \cite{MRT} by
characteristic features (\emph{viz.}, decay constants or
condensates) of light pseudoscalar~mesons \cite{WLr}.

\section{Quark--Antiquark Bound-State Formalism}The homogeneous
Bethe--Salpeter equation describes, in Poincar\'e-covariant
manner, a bound state $|{\rm B}(P)\rangle$ of mass $\widehat M$
and momentum $P,$ formed by two particles of relative momentum
$p,$ by its Bethe--Salpeter amplitude $\Phi(p,P)$. \emph{One
ingredient\/} are the constituents' full propagators, in the case
of spin-$\frac{1}{2}$ fermions $i$ given by mass $M_i(p^2)$ and
wave-function renormalization~$Z_i(p^2)$:$$S_i(p)=\frac{{\rm
i}\,Z_i(p^2)}{\slashed{p}-M_i(p^2)+{\rm i}\,\varepsilon}\
,\qquad\slashed{p}\equiv p^\mu\,\gamma_\mu\
,\qquad\varepsilon\downarrow0\ ,\qquad i=1,2\ .$$The \emph{other
ingredient\/} are the interactions responsible for bound-state
formation. Ignoring their dependence on time components of momenta
\emph{and\/} the above propagators' dependence on zero momentum
components \emph{squared\/} led to a bound-state equation
\cite{WLe} for a Salpeter amplitude~\cite{SE}
$$\phi(\bm{p})\propto\int{\rm d}p_0\,\Phi(p,P)\ ,$$interpretable
as equal-time bound-state wave function. An integral kernel
$K(\bm{p},\bm{q})$ captures the entirety of, by assumption
instantaneous, effective interactions of all bound-state
constituents.\pagebreak

The bound-state equation proposed in Ref.~\cite{WLe}, formulated
in terms of the kinetic energies,
$$E_i(\bm{p})\equiv\sqrt{\bm{p}^2+M_i^2(\bm{p}^2)}\ ,$$and the
energy projectors onto positive and negative energies of the
bound-state constituents~$i,$
$$\Lambda_i^\pm(\bm{p})\equiv\frac{E_i(\bm{p})\pm
\gamma_0\,[\bm{\gamma}\cdot\bm{p}+M_i(\bm{p}^2)]}{2\,E_i(\bm{p})}\
,$$reads, for the case of fermion--antifermion bound states in the
center-of-momentum frame~\cite{WLe},
\begin{align*}\phi(\bm{p})=Z_1(\bm{p}^2)\,Z_2(\bm{p}^2)
\int\frac{{\rm d}^3q}{(2\pi)^3}&\left(\frac{\Lambda_1^+(\bm{p})\,
\gamma_0\,[K(\bm{p},\bm{q})\,\phi(\bm{q})]\,\Lambda_2^-(\bm{p})\,
\gamma_0}{\widehat M-E_1(\bm{p})-E_2(\bm{p})}\right.\\&
\hspace{-1.4ex}-\left.
\frac{\Lambda_1^-(\bm{p})\,\gamma_0\,[K(\bm{p},\bm{q})\,\phi(\bm{q})]\,
\Lambda_2^+(\bm{p})\,\gamma_0}{\widehat M+E_1(\bm{p})+E_2(\bm{p})}
\right).\end{align*}

For any one-particle states $|B(P)\rangle$ normalized according to
the Lorentz-invariant condition$$\langle B(P)|B(P')
\rangle=(2\pi)^3\,2\,P_0\,\delta^{(3)}(\bm{P}-\bm{P}')\ ,$$the
normalization condition for the corresponding Salpeter amplitudes
$\phi(\bm{p})$ is given by \cite{L,RMMP,OVW}$$\int\frac{{\rm
d}^3p}{(2\pi)^3}\,{\rm Tr}\! \left[\phi^\dag(\bm{p})\,
\frac{\gamma_0\,[\bm{\gamma}\cdot\bm{p}+M_1(\bm{p}^2)]}
{E_1(\bm{p})}\,\phi(\bm{p})\right]=2\,P_0\ .$$

\section{Assuming Generalized Flavour Symmetry}Things simplify
considerably if the propagator functions of involved fermion and
antifermion happen to be identical, $M_1(\bm{p}^2)=M_2(\bm{p}^2)$
and $Z_1(\bm{p}^2)=Z_2(\bm{p}^2);$ the Salpeter amplitude of every
pseudoscalar bound state is then fully defined by just two
Lorentz-scalar components,~$\varphi_{1,2}(\bm{p})$:
$$\phi(\bm{p})=\frac{1}{\sqrt{3}}\left[\varphi_1(\bm{p})\,
\frac{\gamma_0\,[\bm{\gamma}\cdot\bm{p}+M(\bm{p}^2)]}{E(\bm{p})}+
\varphi_2(\bm{p})\right]\gamma_5\ .$$

For $K(\bm{p},\bm{q})$ compatible with spherical and rather
specific Fierz symmetries of the effective interactions, our
bound-state equation governing $\phi(\bm{p})$ collapses to an
eigenvalue problem~\cite{WLs} fixing the radial parts
$\varphi_{1,2}(p),$ $p\equiv|\bm{p}|,$ of $\varphi_{1,2}(\bm{p}),$
with a single, spherically symmetric potential $V(r),$
$r\equiv|\bm{x}|,$ encoding the configuration-space interactions
between bound-state constituents:\begin{align}&E(p)\,\varphi_2(p)
+\frac{2\,Z^2(p^2)}{\pi\,p} \int\limits_0^\infty{\rm d}q\,q\,{\rm
d}r\sin(p\,r)\sin(q\,r)\,V(r)\,\varphi_2(q)=\frac{\widehat M}{2}\,
\varphi_1(p)\ ,\nonumber\\&E(p)\,\varphi_1(p)=\frac{\widehat M}{2}
\,\varphi_2(p)\ .\label{c}\end{align}

\begin{figure}[hbt]
\centering\includegraphics[scale=1.89231,clip]{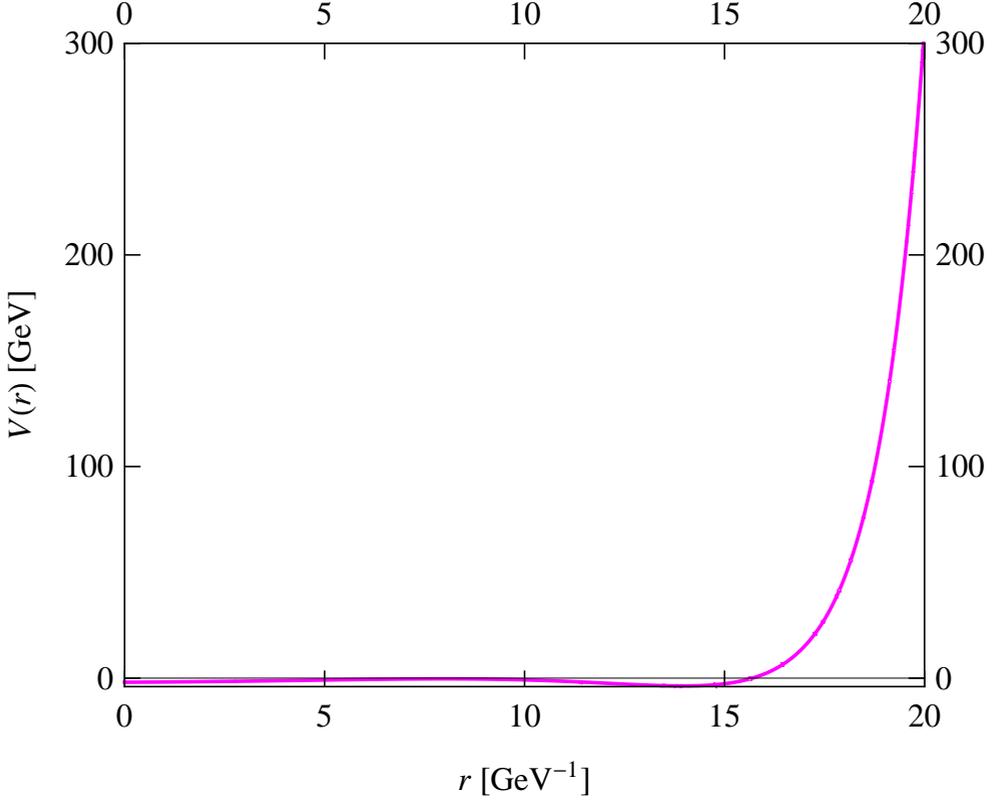}
\caption{Effective interquark central potential $V(r)$ pinned down
by inverting the radial Bethe--Salpeter problem (\ref{c}) for a
Salpeter-amplitude input derived from some quark-propagator model
solution \cite{MT,PM}. The almost flatness of the potential $V(r)$
near the origin in combination with its extraordinarily steep rise
to infinity renders the form of $V(r)$ pretty close but clearly
not identical to that of a square-well~potential.}\label{P}
\end{figure}

For the actual case of interest, mesonic bound states of quarks
and antiquarks, the effective \emph{interaction potential\/}
$V(r)$ was extracted pointwise \cite{WLp,WLpa,WLpb} by
straightforward inversion \cite{WLi,WLia} of our
Bethe--Salpeter-inspired bound-state equation \cite{WLe}, starting
from that Salpeter amplitude $\phi(\bm{p})$ that represents
massless pseudoscalar mesons. The latter, in turn, is connected
\cite{WLc,WLca,WLcb} to the chiral-quark propagator \cite{MT,PM}
by a Ward--Takahashi identity of QCD \cite{MRT}. The emerging
confining $V(r)$ rises, from its slightly negative value at $r=0,$
rather steeply to infinity (Fig.~\ref{P}).

\section{Pseudoscalar-Meson Properties Revisited}In order to
scrutinize pseudoscalar quark--antiquark bound states for physical
(\emph{i.e.}, non-chiral) quark masses, we ought to find the
corresponding solutions to our set (\ref{c}) of coupled equations
for the two independent radial Salpeter components
$\varphi_{1,2}(p).$ Our task is greatly facilitated by a
definitely obvious move enabled by the purely algebraic nature of
one of the two relations~(\ref{c}): inserting any of the two
relations (1) into the other leads to single explicit eigenvalue
problems for either $\varphi_1(p)$ or $\varphi_2(p)$ with
mass-squared eigenvalue $\widehat M^2$ \cite{WLp,WLpa,WLpb,WLs};
conversion to equivalent matrix eigenvalue problems by expansion
over a convenient function-space basis is among the standard
solution procedures \cite{WLm,WLma,WLmb,WLmc,WLf}. With these
solutions at hand, we may create trust in the reliability of our
approach by assessing and exploring its predictions for hadronic
observables.

The spatial extension of the pion deduced, from the numerical
ground-state solution to our bound-state formalism inferred along
the above lines, in form of the pion's average interquark distance
$\langle r\rangle=0.478\;\mbox{fm}$ or root-mean-square radius
$\sqrt{\langle r^2\rangle}=0.529\;\mbox{fm}$ fits nicely to the
pion's electromagnetic charge radius, $\sqrt{\langle
r_\pi^2\rangle}=(0.672\pm0.008)\;\mbox{fm}$ \cite{PDG}.
Nevertheless, this agreement won't qualify as confirmation of the
credibility of our framework since the Salpeter amplitude for
chiral quarks served already as input to the inversion procedure
that generated the shape of our potential $V(r),$ and the
experimental $u/d$ quark masses are pretty close to their
chiral~limit.

Consequently, we have to look for different and more significant
criteria which allow us to appraise the reasonableness of any
inversion considerations. Fortunately, this proves to be not too
difficult: Equating the residues of pseudoscalar-meson pole terms
in both axial-vector and pseudoscalar vertex functions entering
into the axial-vector Ward--Takahashi identity of QCD expressing
the invariance of QCD under chiral transformations leads to a
generalization \cite{MRT} of the Gell-Mann--Oakes--Renner relation
\cite{GOR}; this innovation relates, for a pseudoscalar~bound
state $|B(P)\rangle,$ its (weak) \emph{decay constant\/} $f_B,$
defined in terms of the axial-vector quark current~by$$\langle0|
{:\!\bar\psi_1(0)\,\gamma_\mu\,\gamma_5\,\psi_2(0)\!:}|B(P)\rangle
={\rm i}\,f_B\,P_\mu\ ,$$--- which, consequently, may be found by
projection of $\phi(\bm{p})$ onto the axial-vector
current,~\emph{i.e.},$$f_B\propto\int{\rm d}^3p\,{\rm
Tr}[\gamma_0\,\gamma_5\,\phi(\bm{p})]$$--- and its \emph{in-hadron
condensate\/} \cite{MRT} (universalizing the notion of quark
vacuum condensates)\footnote{A recent, comprehensive,
Bethe--Salpeter-rooted evaluation of in-hadron condensates may be
found in Ref.~\cite{HGKL}.} $${\mathbb C}_B\equiv
\langle0|{:\!\bar\psi_1(0)\,\gamma_5\,\psi_2(0)\!:}|B(P)\rangle
\propto\int{\rm d}^3p\,{\rm Tr}[\gamma_5\,\phi(\bm{p})]$$to the
mass $\widehat M_B$ of this pseudoscalar bound state and the two
relevant quark mass parameters~in the QCD Lagrangian \cite{MRT}.
For the simpler case of equal quark masses $m,$ this relationship
reads\begin{equation}f_B\,\widehat M_B^2=2\,m\,{\mathbb C}_B\
.\label{GOR}\end{equation}

Compatibility with Eq.~(\ref{GOR}) may be inspected by solving our
formalism with~the~previously established potential $V(r)$ for
bound states of chiral, $u/d,$ and $s$ quarks, taking
advantage~of~the appropriate model propagator functions
\cite{MT,PM}. Comparison of the quark masses $m,$~fixed~by the
thus determined values of $\widehat M_B,$ $f_B,$ and ${\mathbb
C}_B,$ with the current-quark masses $\overline{m}(\mu)$ in
modified minimal subtraction at scale $\mu$ proves that our
findings for $m$ are in the right ballpark~(Table~\ref{F}).

\begin{table}[hbt]\centering\caption{Predictions of the
Bethe--Salpeter-inspired bound-state equation (\ref{c}), with
effective interaction potential $V(r)$ as depicted in
Fig.~\ref{P}, for masses $\widehat M_B,$ decay constants $f_B$ and
in-meson condensates ${\mathbb C}_B$ of the lightest pseudoscalar
mesons, and confrontation of the quark-mass parameters $m$
resulting from the more general Gell-Mann--Oakes--Renner relation
(\ref{GOR}) with their minimal-subtraction counterparts
$\overline{m}(\mu).$}\label{F}\begin{tabular}{rrccd{1.4}c}
\toprule\multicolumn{1}{c}{Constituents}&\multicolumn{1}{c}
{$\widehat M_B$}&$f_B$&${\mathbb C}_B$&\multicolumn{1}{c}{$m$}&\ \
$\overline{m}(2\;\mbox{GeV})$\\&\multicolumn{1}{c}{\ \
$[\mbox{MeV}]$\ \ }&\ \ $[\mbox{MeV}]$\ \ &\ \ $[\mbox{GeV}^2]$\ \
&\multicolumn{1}{c}{\ \ $[\mbox{MeV}]$\ \ }&\ \ $[\mbox{MeV}]$
\cite{PDG}\\\midrule chiral quarks&6.8&151&
0.585&0.0059&---\\[1ex]$u$/$d$ quarks&148.6&155&0.598&2.85&
$3.5^{+0.7}_{-0.3}$\\[1ex]$s$ quarks&620.7&211&0.799&51.0&
$96^{+8}_{-4}$\\\bottomrule\end{tabular}\end{table}

\section{Summary of Findings and Conclusions}In order to establish
whether or not it is justified to lend trust to the outcomes of an
approach to bound states proposed some time ago \cite{WLe} and
residing, as far as its compatibility with Poincar\'e covariance
is concerned, somewhere in the vast range between bound-state
descriptions along the ideas of Salpeter and Bethe, on the one
hand, and static approximations, on the other hand, we evaluated
the performance of the investigated framework's predictions for
those properties of the lightest pseudoscalar mesons that happen
to be related by an advancement of the insight gained by
Gell-Mann, Oakes, and Renner \cite{GOR}: In brief, the formalism
of Ref.~\cite{WLe} is still alive.

\end{document}